\documentclass[slac_one]{revtex4}
\usepackage{graphicx}
\usepackage{fancyhdr}
\usepackage{wrapfig,rotating}

\begin{document}

\title{Beyond the Standard Model Higgs Boson Searches at the Tevatron} 

%

\author{Tim Scanlon on behalf of the D0 and CDF Collaborations}
\affiliation{Imperial College London, Physics
Department, Prince Consort Road, London SW7 2AZ, United Kingdom.}

\begin{abstract}
Results are presented for beyond the Standard
Model Higgs boson searches using up to 4.2~fb$^{-1}$ of data from
Run II at the Tevatron. No significant excess is observed in any of
the channels so 95\% confidence level limits are presented.
\end{abstract}

\maketitle

\thispagestyle{fancy}


\section{Introduction}

The search for the Higgs boson is one of the main goals in High Energy Physics and one of the highest priorities at Run II of the Tevatron. There are many alternative
Higgs boson models beyond the SM, including Supersymmetry
(SUSY)~\cite{mssm} and Fermiophobic Higgs bosons~\cite{haber}, which can
actively be probed at the Tevatron, and in the absence of an excess
constrained. The latest limits for several SUSY searches are presented
in Section~\ref{sec:mssm} and for the Fermiophobic Higgs boson searches in
Section~\ref{sec:fermiophobic}. More information on all
these searches, along with the latest results, can be found on the CDF and D0 public results webpages~\cite{cdf_web_page,dzero_web_page}.

\section{Minimal Supersymmetric Standard Model Higgs Boson Searches} \label{sec:mssm}

The Minimal Supersymmetric extension of the SM (MSSM) \cite{mssm}
introduces two Higgs doublets which results in five physical Higgs
bosons after electroweak symmetry breaking. Three of the Higgs bosons are
neutral, the CP-odd scalar, $A$, and the CP-even scalars, $h$ and
$H$ ($h$ is the lighter and SM like), and two are
charged, $H^{\pm}$.

At tree level only two free parameters are needed for all couplings
and masses to be calculated. These are chosen as the mass of the
CP-odd scalar ($m_{A}$) and tan$\beta$, the ratio of the two vacuum
expectation values of the Higgs doublets.

The Higgs boson production cross section in the MSSM is proportional to
the square of tan$\beta$. Large values of tan$\beta$ thus result in
significantly increased production cross sections compared to the
SM.  Moreover, one of the CP-even scalars and the CP-odd scalar are
degenerate in mass, leading to a further approximate doubling of the
cross section.

The main production mechanisms for the neutral Higgs bosons are the
$gg, b\bar{b} \rightarrow\phi$ and $gg, q\bar{q} \rightarrow \phi +
b\bar{b}$ processes, where $\phi = h, H, A$. The branching ratio of
$\phi \rightarrow b\bar{b}$ is around 90\% and $\phi \rightarrow
\tau^{+}\tau^{-}$ is around 10\%. This results in three channels of
interest: $\phi \rightarrow \tau^{+}\tau^{-}$, $\phi b\rightarrow
b\bar{b}b$ and $\phi b\rightarrow \tau^{+}\tau^{-}b$. The overall
experimental sensitivity of the three channels is similar due to the
lower background from the more unique signature of the $\tau$
decays.

\subsection{Higgs $\rightarrow \tau^{+}\tau^{-}$} \label{tautau}



D0's most recent search is in the $\tau_{\mu}\tau_{had}$ final state
using 1.2 fb$^{-1}$ of Run II data, where $\tau_{had}$ refers to a
hadronic decay and $\tau_{\mu}$ to a leptonic decay (to a $\mu$) of the $\tau$. This result is an extension to, and combined with, the published
1~fb$^{-1}$ result which also included the $\tau_{\mu}\tau_{e}$,
and $\tau_{e}\tau_{had}$ channels~\cite{p17tautau}. CDF have published a search combining the $\tau_{\mu}\tau_{e}$,
$\tau_{\mu}\tau_{had}$ and $\tau_{e}\tau_{had}$ final states using
1.8~fb$^{-1}$ of RunII data~\cite{cdftautau}.

Both searches require events to have an isolated $\mu$ ($e$), separated from an opposite signed $\tau_{had}$ (or $e$ for the $\tau_{\mu}\tau_{e}$ channel). Hadronic $\tau$
candidates are identified at D0 by the standard neural
networks designed to distinguish $\tau_{had}$ from multi-jet events and at CDF by using a variable size isolation cone. To
minimise the $W$+jets background events are removed which have a large
$W$ transverse mass (D0) or by placing a cut on the relative
direction of the visible $\tau$ decay products and the missing
$E_{T}$ (CDF).

\begin{figure}[htb]
\centering
\includegraphics[width=0.5\columnwidth]{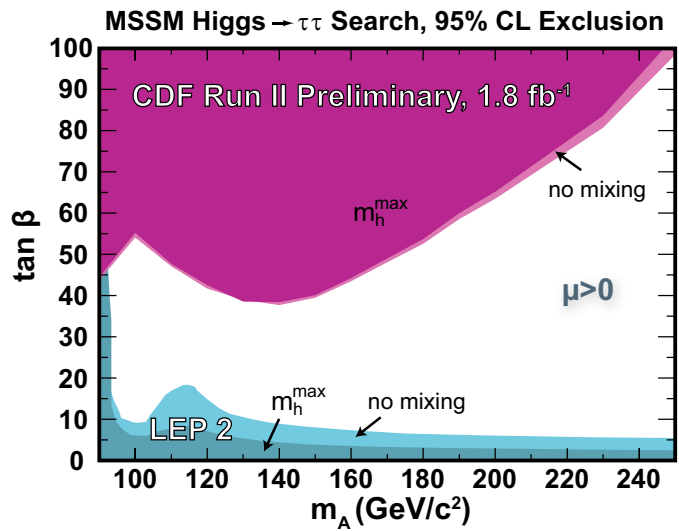}
\caption{The 95\% CL excluded region in the tan$\beta$ versus m$_{a}$ plane from the CDF Higgs $\rightarrow \tau^{+}\tau^{-}$ analysis for $\mu > 0$ in the no-mixing and
$m_{h}^{max}$ scenarios~\cite{scenarios}. The shaded blue area is the region excluded
by LEP~\cite{lep}.}
\label{Fig:tautaulimits}
\end{figure}

In both analyses the dominant $Z/\gamma \rightarrow \tau\tau$ background is modelled using PYTHIA~\cite{pythia} and the multi-jet/$W$+jets contribution is modelled using data. Limits are set using the visible mass distribution ($m_{vis}$),
which is the invariant mass of the visible $\tau$ products and the
missing $E_{T}$. The 95\% confidence level (CL) limit is shown in
Fig.~\ref{Fig:tautaulimits} for the CDF search interpreted in one of the
standard MSSM scenarios~\cite{scenarios}.

\subsection{Higgs $+~b\rightarrow b\bar{b}b$}


This channel has a signature of at least three $b$ jets, with the background consequentially dominated by heavy flavour multi-jet
events. D0 and CDF have both conducted searches in this
channel, using 2.6~fb$^{-1}$ and 2.2~fb$^{-1}$ of data respectively.

Both searches require three $b$-tagged jets, D0 uses its
standard neural network $b$-tagging algorithm~\cite{btagthesis} and CDF uses its standard secondary vertex algorithm. Due to the difficultly of simulating the heavy flavour multi-jet background, both analyses use data driven approaches. D0 uses a fit to data over several different $b$-tagging criteria whereas CDF uses fits to dijet invariant and secondary vertex mass templates to determine the heavy flavour sample composition.

\begin{figure}[htb]
\centering
\includegraphics[width=0.55\columnwidth]{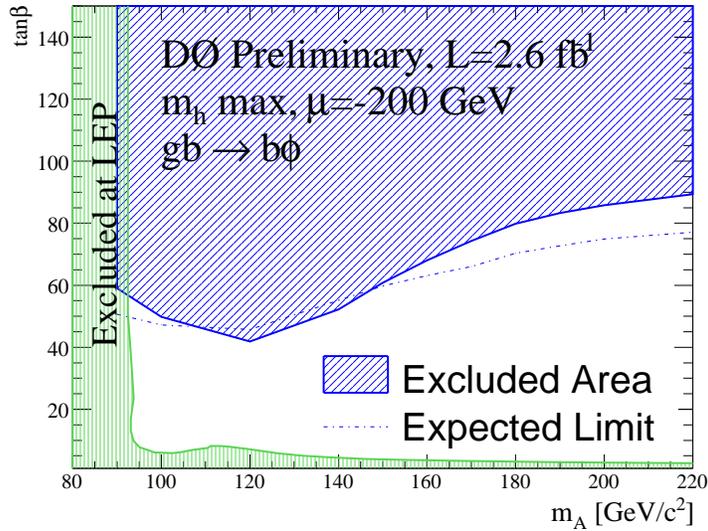}
\caption{The 95\% CL excluded region in the tan$\beta$ versus m$_{a}$ plane from the D0 Higgs $+~b\rightarrow b\bar{b}b$ analysis for the $\mu<0$, $m_h^{max}$ scenario~\cite{scenarios}. The green area is the region excluded by LEP~\cite{lep}.}
\label{Fig:hbblimits}
\end{figure}

To increase the sensitivity of the analysis, D0 splits it into exclusive three,
four and five-jet channels, training a likelihood to distinguish the signal from background in each. Limits are set by both CDF and D0 on the Higgs boson production cross-section
times branching ratio using the dijet invariant mass as the discriminating
variable. The limits are then interpreted in the standard scenarios, as shown for D0 in Fig.~\ref{Fig:hbblimits}.

\subsection{Higgs $+~b\rightarrow \tau^{+}\tau^{-}b$} \label{btautau}


D0 has performed a search for the $\tau_{\mu}\tau_{had}$ signature using 4.3~fb$^{-1}$ of Run II data. Events are selected by requiring
an isolated $\mu$ separated from an opposite sign
$\tau_{had}$ candidate, along with a $b$ tagged jet. The $\tau_{had}$ decays are
identified using the standard D0 neural networks and $b$ jets using the neural network
$b$-tagging algorithm. The dominant backgrounds are $t\bar{t}$, $W$+jets, multi-jet and $Z$+jet events. The
multi-jet and $W$+jets backgrounds are estimated from data with $t\bar{t}$
modelled using ALPGEN~\cite{alpgen} interfaced with PYTHIA.

\begin{figure}[htbp]
\centering
\hspace{1cm}
\includegraphics[width=0.6\columnwidth]{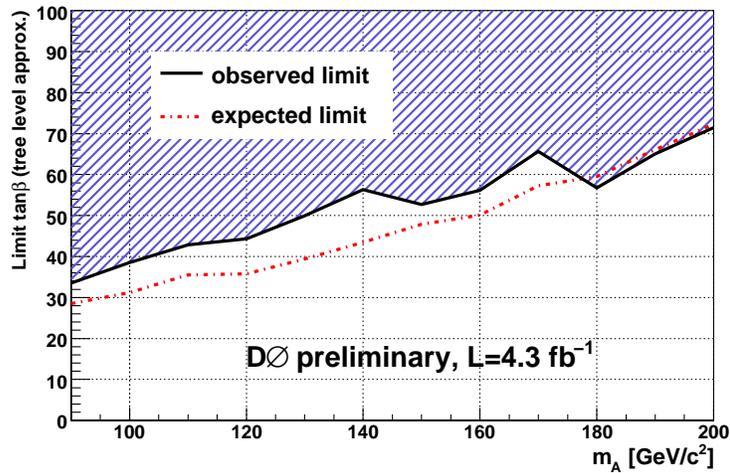}
\caption{The 95\% CL excluded region in the tan$\beta$ versus m$_{a}$ plane from the D0 Higgs $+~b\rightarrow \tau^{+}\tau^{-}b$ analysis for the tree level approximation.}
\label{Fig:btautaulimits}
\end{figure}

To improve the sensitivity of the analysis three discriminants are formed which differentiate the signal from the $t\bar{t}$, multi-jet and $Z$+light parton events respectively. The three discriminants are combined to form the final discriminant which is used to set limits.
Figure~\ref{Fig:btautaulimits} shows the resulting 95\% CL exclusion
in the tan$\beta - m_{A}$ plane, which is the best limit to date from a single channel at a hadron collider.

\subsection{Combined Limits}

\begin{figure}[htbp]
\centering
\includegraphics[width=0.48\columnwidth]{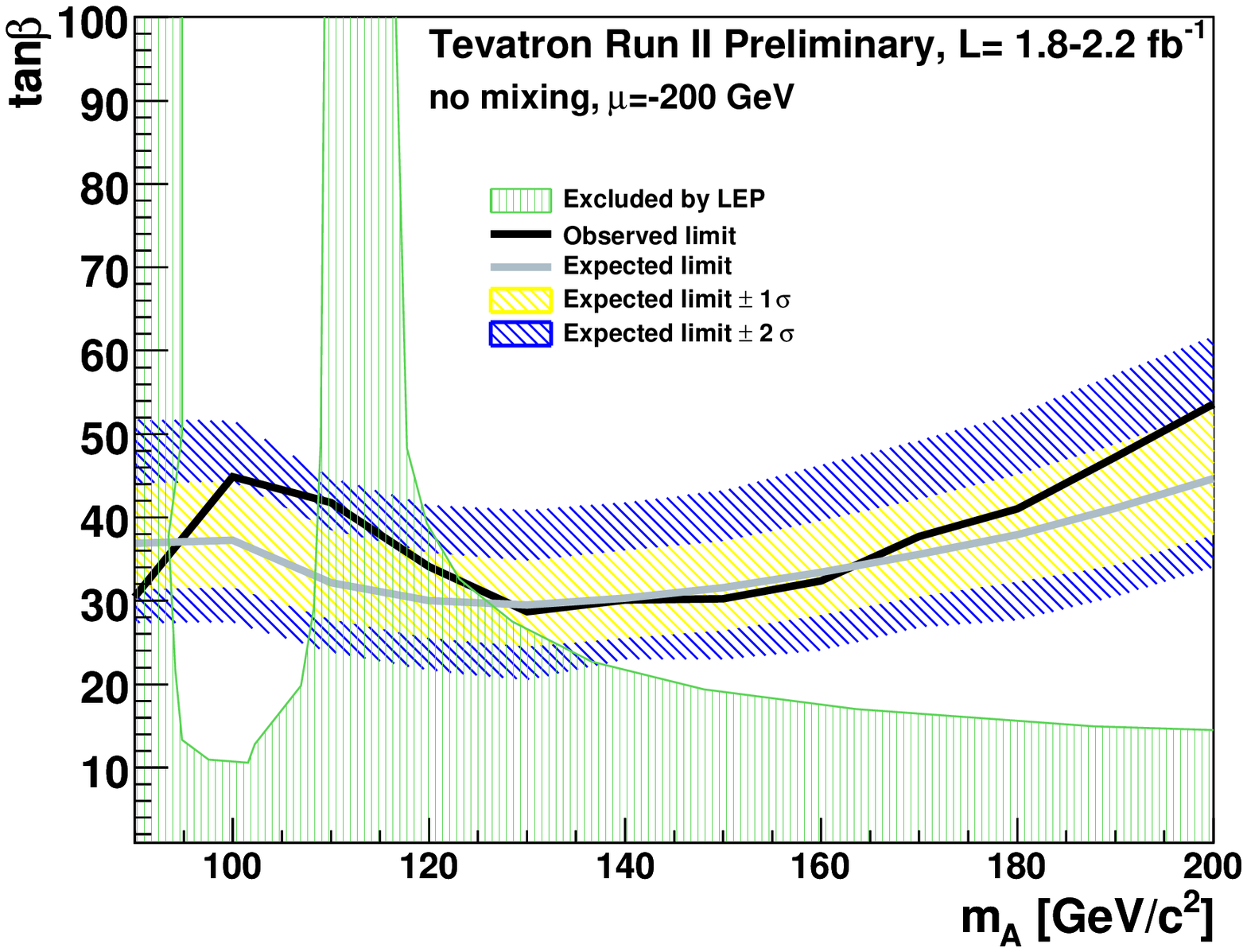}
\includegraphics[width=0.48\columnwidth]{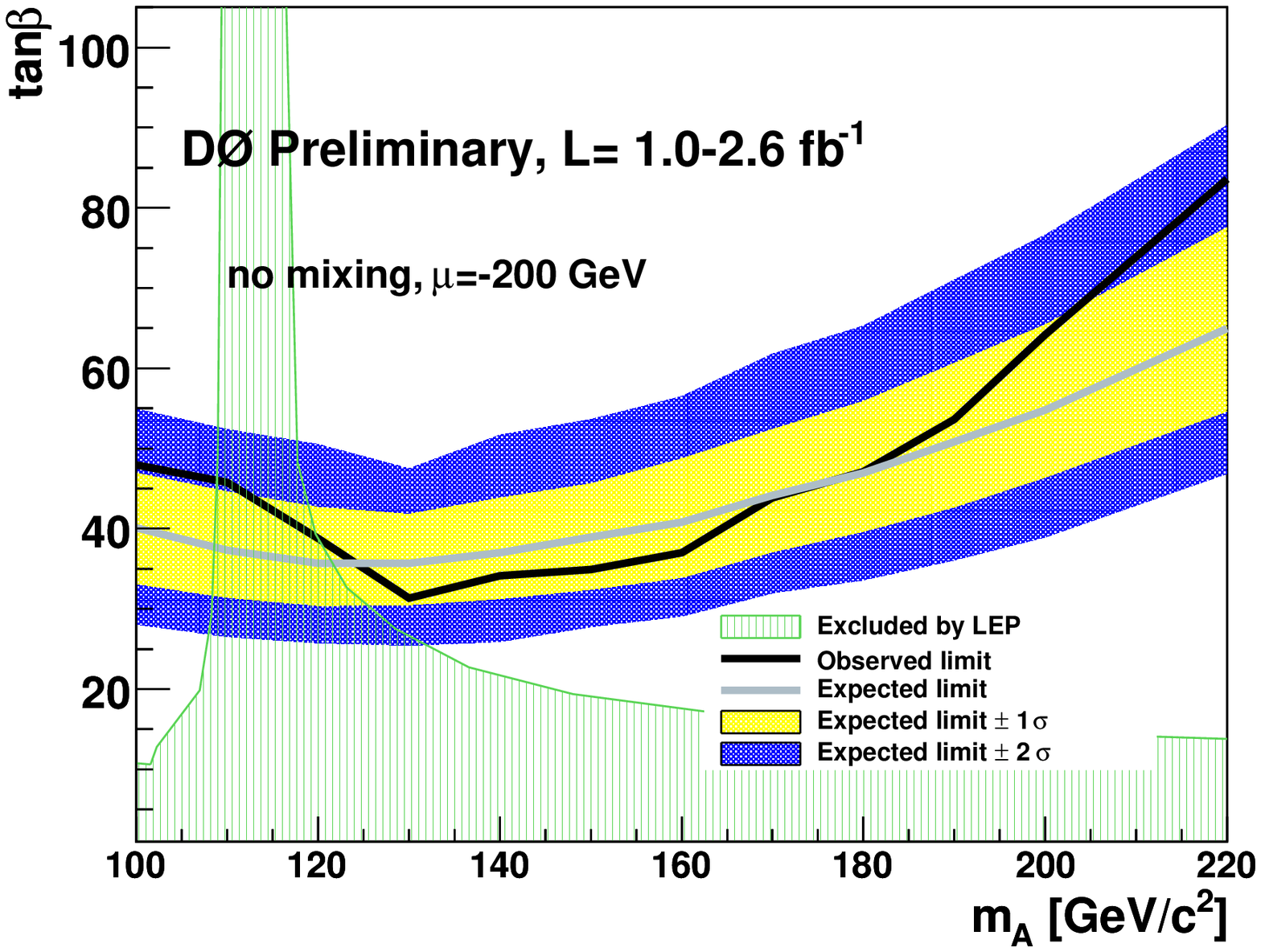}
\caption{The combined D0 (left) and Tevatron (right) 95\% CL excluded regions in the tan$\beta$ versus m$_{a}$ plane for the $\mu<0$, no mixing scenario~\cite{scenarios}. The green area is the region excluded by LEP~\cite{lep}.}
\label{Fig:tevcomblimits}
\end{figure}

The channels described in Sections~\ref{tautau}--\ref{btautau} are complementary and can be combined to increase the reach of the MSSM Higgs boson searches at the Tevatron. D0 has combined its three neutral Higgs boson channels (using an earlier version of the $\tau^{+}\tau^{-}b$ analysis based on only 1.2~fb$^{-1}$ of RunII data) and interpreted the limits in the standard MSSM scenarios. A combined Tevatron limit on the MSSM Higgs sector has also been produced from D0 and CDF's Higgs $\rightarrow \tau^{+}\tau^{-}$ channels. The combined Higgs $\rightarrow \tau^{+}\tau^{-}$ result has been interpreted in a quasi-model independent limit, as well as in the standard scenarios. Both the D0 and Tevatron combined limits are shown in Fig.~\ref{Fig:tevcomblimits} for a typical scenario. The combined limits are the most stringent to date at a hadron collider in the tan$\beta - m_{A}$ plane.

\subsection{Charged Higgs Boson Searches}



\begin{figure}[htb]
\centering
\includegraphics[width=0.5\columnwidth]{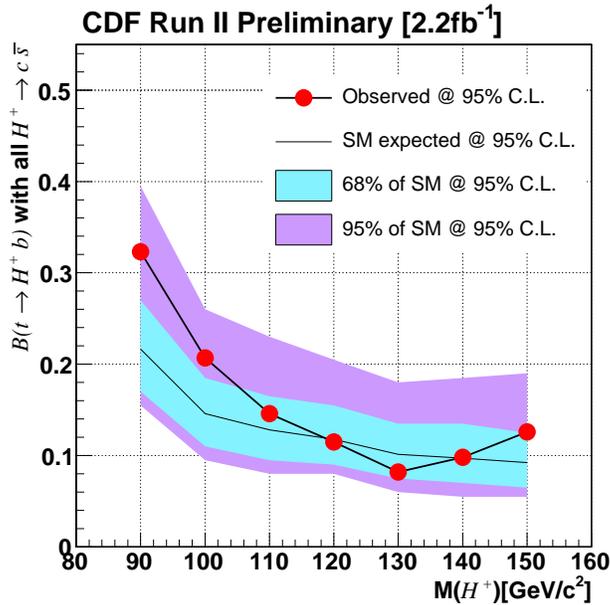}
\caption{CDF 95\% CL upper limits on the branching ratio of $t\rightarrow H^{+}b$.}
\label{Fig:chargedHiggsCDF}
\end{figure}

CDF has published a search for $H^{+}$ using 2.2~fb$^{-1}$ of data
in the low tan$\beta$ region, where the charged Higgs boson predominantly decays
as $H^{+}\rightarrow c\bar{s}$~\cite{cdfchargedhiggs}, in $t\bar{t}$ decays. Events are required to have one
isolated lepton, missing $E_{T}$ and four jets, two of which must be tagged as $b$ jets by the
secondary vertex tagging algorithm. The two non $b$-tagged jets are
assumed to be from the $W^{+}$ or $H^{+}$ decay. Mass templates for
$H^{+}$ and $W^{+}$ are fit to the data dijet mass distribution to
extract uppper limits on the branching ratio of $t \rightarrow H^{+}
b$. The dominant background is $t\bar{t}$ which is modelled using
PYTHIA. The 95\% CL upper limits on the B($t \rightarrow H^{+}b$) are shown in
Fig.~\ref{Fig:chargedHiggsCDF}.

D0 carried out a search using 1 fb$^{-1}$ of data~\cite{d0chargedhiggs}. The search was
conducted in the lepton+jets ($e$+jets, $\mu$+jets) and dilepton
($ee$, $\mu\mu$, $e\mu$, $\mu\tau$, $e\tau$) decay channels.
Searches were conducted in the tan$\beta < 1$ region where the
tauonic decay, $H^{+}\rightarrow \tau \nu$, dominates and in the
tan$\beta > 1$ region where the leptophobic decay, $H^{+}\rightarrow
c\bar{s}$, dominates. In the leptophobic decay a suppression of the
expected $t\bar{t}$ yield would indicate the presence of the $H^{+}$
decay in all channels. In the tauonic decay a deficit in all
channels except for the lepton+jets channel would indicate the
presence of the $H^{+}$. The signal was modelled using PYTHIA and
$t\bar{t}$ production was modelled using ALPGEN interfaced with PYTHIA.

\subsection{Next-to-MSSM Higgs Bosons Searches}


In the next-to-MSSM (nMSSM)~\cite{nmssm} the branching ratio of
Higgs$\rightarrow b\bar{b}$ is greatly reduced. Instead the Higgs
boson predominantly decays to a pair of lighter neutral pseudoscalor Higgs
bosons, $a$. The nMSSM scheme is interesting as it allows the LEP
limit on the $h$ boson to be naturally lowered to the general Higgs
boson search limit from LEP of $M_{h} > 82$~GeV~\cite{lepgeneral}.

\begin{figure}[htb]
\centering
\includegraphics[width=0.5\columnwidth]{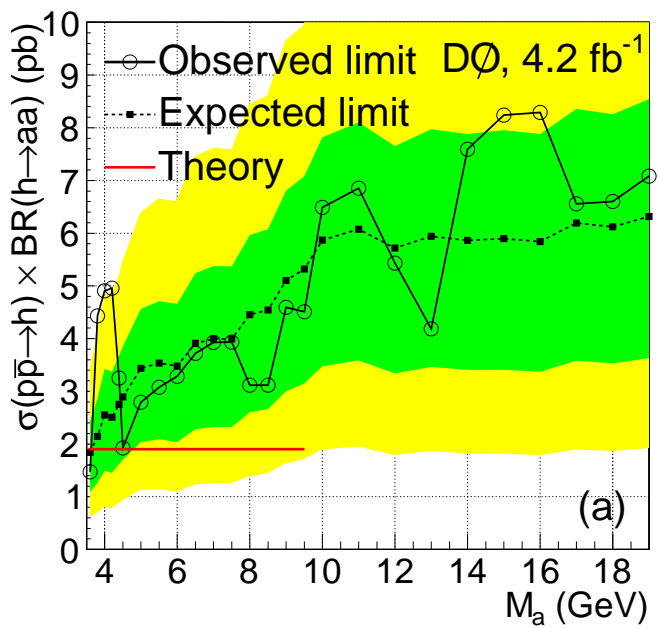}
\caption{D0 95\% CL limits for $\sigma(p\bar{p}\rightarrow h + X)\rightarrow
aa$, assuming $M_{h}$ = 100~GeV. The solid red line shows the signal for BR($h\rightarrow aa =1$).}
\label{Fig:amumutautau}
\end{figure}

D0 has conducted two nMSSM searches \cite{D0nmssm}. Firstly for the case where $a$ predominantly decays to two muons resulting in an event signature of two collinear muons pairs. Due to the extreme collinearity of the
$\mu$ pair the event selection requires two well separated muons each matched to a companion track. The main backgrounds are multi-jet events
and $Z/\gamma \rightarrow \mu\mu + jets$. The multi-jet events are modelled
using a data control region and the $Z/\gamma$ background is modelled
using PYTHIA. Limits are set by counting events in a 2D window around the two companion
track-$\mu$ invariant mass peaks. Secondly, for the case where the $a$ predominantly decays to two $\tau$ particles. Due to the lack of an
observed resonance and the difficulty of triggering on a four $\tau$
signature, one of the $a$ bosons is required to decay to two muons.
The event selection requires an isolated pair of muons and significant missing $E_{T}$ opposite the $\mu$ pair. The multi-jet background is modelled using a data control
region and other electroweak and top backgrounds are modelled using
PYTHIA. Limits are set by counting events in a mass window around the di-$\mu$ invariant
mass peak and are shown in Fig.~\ref{Fig:amumutautau}.

CDF has also conducted a search for a light nMSSM Higgs boson using 2.7~fb$^{-1}$ of data in top quark decays, where $t\rightarrow W^{\pm(*)}ab$ and $a\rightarrow\tau\tau$. The $\tau$ particles are identified by the presence of additional isolated tracks in the event due to their low $p_{T}$. The dominant background is from soft parton interactions and is modelled using data. Limits are set on the branching ratio of a top quark decaying to a charged Higgs boson from a fit to the $p_{T}$ spectrum of the lead isolated track.

\section{Fermiophobic Higgs Boson Searches}
\label{sec:fermiophobic}

The Standard Model Higgs boson branching ratio to a pair of
photons is small. There are however several models where the decay
of the Higgs boson to fermions is suppressed. In these models the
decay of the Higgs boson to photons is greatly enhanced. Both D0
and CDF \cite{cdffermiphobic} have carried out searches for the Fermiophobic Higgs boson
using 4.2 and 3.0 fb$^{-1}$ respectively.

D0 requires two photon candidates in the central calorimeter, with jets misidentified as photons rejected by use of a neutral network. Electrons are suppressed by requiring
that the photon candidates are not matched to activity in the
tracking detectors. The three main background sources are estimated
separately: the jet and diphoton backgrounds are estimated from data
and the Drell-Yan contribution is estimated using PYTHIA.

CDF's search also requires two photons, with only one of them required to be in the central region of the calorimeter. This looser photon requirement approximately doubles the acceptance compared to requiring both photons in the central region. In addition a cut is placed on the transverse
momentum of the two photons which significantly reduces the background, which is estimated using
a purely data-based approach.

\begin{figure}[htb]
\centering
\includegraphics[width=0.5\columnwidth]{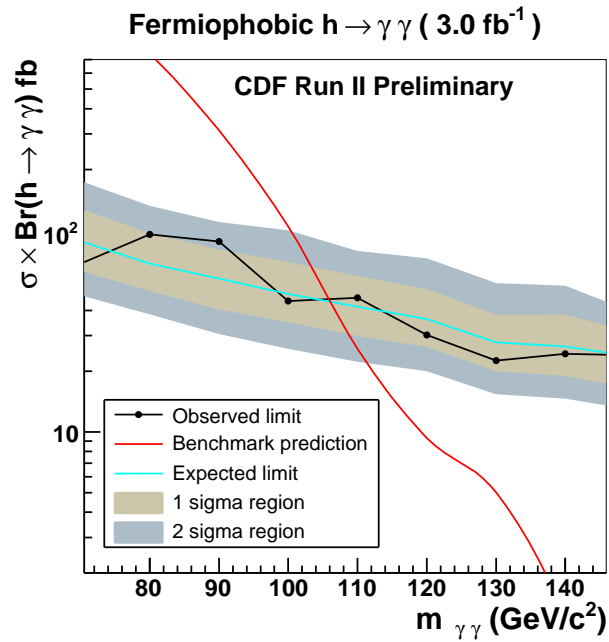}
\caption{CDF 95\% CL limits on $\sigma \times BR$ as a function of the
Fermiophobic Higgs boson mass.}
\label{Fig:fermiophobichiggs}
\end{figure}

Upper limits are set on the Higgs boson production cross section times
branching ratio using the diphoton mass as the discriminating
variable. The 95\% CL limits are shown in Fig.~\ref{Fig:fermiophobichiggs}
for the CDF search.


\section{Conclusions}

CDF and D0 have a wide variety of beyond the Standard Model
Higgs boson searches, presented here using up to 4.2~fb$^{-1}$ of data. These
searches are already powerful, and have set some of the best limits in
the world. No signal has been observed yet, but with their rapidly improving sensitivity, due to both improved analysis techniques and the addition of between 2--5 times more data (which has already been recorded), these analyses will continue to probe extremely interesting regions of parameter space, promising many exciting results in the near future.

\begin{acknowledgements}
I would like to thank all the staff at Fermilab, the Tevatron accelerator division along with the CDF and the D0 Collaborations.
\end{acknowledgements}

\end{document}